\documentclass[suppldata]{interact}

\usepackage{epstopdf}
\usepackage[caption=false]{subfig}

\usepackage[longnamesfirst,sort]{natbib}
\bibpunct[, ]{(}{)}{;}{a}{,}{,}
\renewcommand\bibfont{\fontsize{10}{12}\selectfont}

\begin{document}


\title{Play Music: An HCI Oriented Evaluation of Google’s Default Music Player Interface}

\author{
\name{Venkatesh Vijaykumar\textsuperscript{a}\thanks{CONTACT Venkatesh Vijaykumar. Email: vvijaykumar3@gatech.edu}}
\affil{\textsuperscript{a}College of Computing, Georgia Institute of Technology, Atlanta, GA, USA}
}

\maketitle

\begin{abstract}
The work embodied in this paper attempts to suggest a few improvements to the playlist creation task interface of the Google Play Music Android application based on recommended practices encountered in the Human-Computer Interaction discipline. The improvements are largely centered on intuitive navigation and selection actions, in order to facilitate a smoother experience in creating, ordering, and adding to music playlists. The work records the efforts in need-finding, design brainstorming, and prototype design and evaluation. The work was undertaken over a single design life cycle, and is an attempt at applying recommended practices in HCI to a widely used real world application.
\end{abstract}

\begin{keywords}
HCI; UI; UX; prototyping; wireframe; evaluation; music; mobile; Android; interface
\end{keywords}

\section{Introduction}

The task of playlist creation within the Google Play Music application environment, is a cumbersome experience. The problems associated with this task in the interface stem largely from non-intuitive design. There is a lack of intuitive affordances such as a long-press action for selection, or a multiple select function. The final playlist also appears as an unordered list, following the sequence of addition of tracks, with no scope to modify the order. The environment surrounding the interface problem is rather varied, since a music player app can be used in many contexts. However, the basic form of use remains similar. The problems defined are not only particularly challenging for novices but also make for a cumbersome experience for advanced users. The study conducted here attempts at some improvements and re-design approaches to mitigate the problems associated with this task. The subsequent sections delve into the process undertaken to mitigate the issues outlined, and the results obtained and inferences drawn.

\section{Methodology}
\subsection{Need-finding}
\subsubsection{Naturalistic Observation}
The core objective was to observe users while they used the application. The particular observation area would be the task of creation of or addition to a playlist, or performing the same action over multiple tracks. The observations were made on the ease of performing the tasks in order to understand if the user finds it cumbersome to add multiple songs from different artists to a playlist without the presence of a multiple selection and execution function. The context of observation would be in general instances that warrant the use of a music player application. Since the task is not associated with any particular physical activity, the observations can be made ex-situ while remaining faithful to common physical constraints experienced in real world use. The data collected in this step will be largely qualitative and subjective.
This observation exercise was carried out to understand the underlying need for playlists and song queues. In a public transit scenario, playlists or queues were needed intermittently, based on the crowd density, and the ease of cycling through tracks. In an exercise setting, playlists were far more necessary as it was important to not lose focus on the exercise in lieu for switching tracks. In a relaxed scenario, when focused only on listening to music, playlists were seldom a necessity, since the user spends their entire attention on the music and can change tracks or settings as they deem requisite. I guarded against possible observer bias by having a second observer unaware of the study and compared notes for possible instances of bias.

\subsubsection{Participant Observation}
As a participant observer, the author used the application as the primary music player app for a designated period of time. Playlists were created from scratch, and tracks were added to existing lists. The data collected was quantitative in time taken to perform these tasks. Qualitative data in the form of comments on the interface intuitiveness was recorded.
The steps taken in order to perform a participant observation were to:
\begin{itemize}
  \item Make regular use of the application concerned, particularly the task concerned.
  \item Add songs to an existing playlist comprising of at least 10 songs from different artists.
  \item Play the playlist and note the order of play, and determine if it follows a desirable order.
\end{itemize}
The data collected in this plan was a mix of qualitative and quantitative data. It comprised of qualitative data in the form of feedback regarding interface and its functionality. The quantitative part was comprised of timing data for the actions required to complete a task. The application was used for more than a week and in a number of scenarios. Playlists were made for scenarios such as driving and running. The ease of creating a playlist and adding songs to it was objectively noted. The creation of playlists, and addition of tracks to existing lists were timed to gather quantitative insight. Potential bias in data gathering was guarded against by taking extensive notes about the usage of the application in general not just the task concerned.
\subsection{Prototyping}
\subsubsection{Prototype 1}
The first prototype is a card prototype of a redesign that contains most features from the existing application in order to retain the familiarity of the interface. There are added functions and features supporting multiple track selection, addition to new or existing lists, and a global create button on the Playlists tab, which aid the fulfillment of the task of creating and adding to playlists with ease. This prototype caters to the novice users by retaining the original design while adding new functionality.
\begin{figure}[hbt!]
\centering
\resizebox*{12cm}{15cm}{\includegraphics{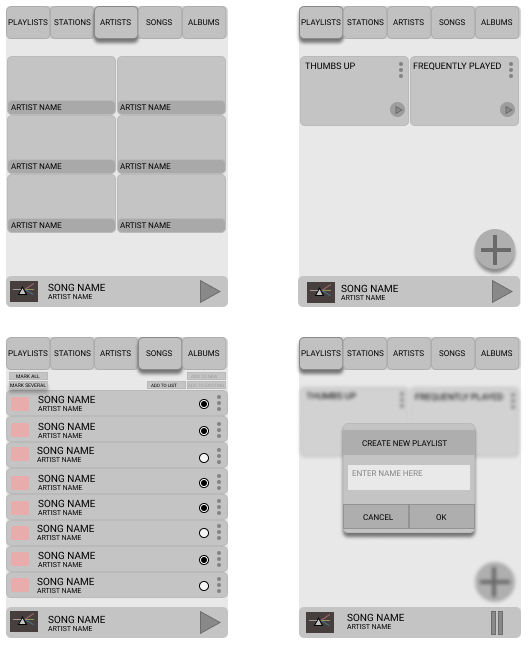}}\hspace{5pt}
\caption{The card prototype. Sequentially displayed clockwise from top-left} \label{card-prototype}
\end{figure}
\newpage
\subsubsection{Prototype 2}
The second prototype is a text based prototype for a gesture heavy interface redesign. Figure 2. explains the design ideas in detail.
This prototype is targeted at the intermediate to advanced user groups. The screen gestures are based on actions that are widely used in the Android operating system. However, the novice user too can find comfort in the interface since the functions are meant to be discoverable, and are not governed by screen gestures alone. The interface should retain the traditional non-gesture based functionality too, in order to cater to as wide an audience as possible.
\begin{figure}[hbt!]
\centering
\resizebox*{7cm}{7cm}{\includegraphics{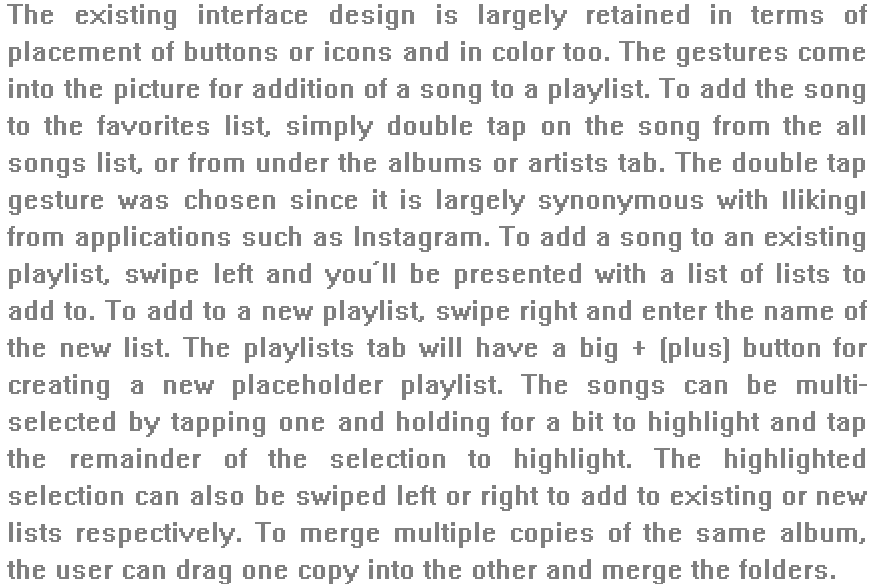}}\hspace{5pt}
\caption{The text prototype scirpt.} \label{txt-prototype}
\end{figure}
\subsubsection{Prototype 3}
The third and final prototype is a Wizard of Oz prototype that is centered about the voice commands based interface redesign proposed in the brainstorming exercise.
The subject is informed of the voice commands as follows:
\begin{itemize}
\item Pause: to pause a playing song
\item Play: to restart a paused song
\item Like: to add currently playing song to favorites
\item Add: to add playing songs to existing list (presented with list names and corresponding option numbers)
\item Add new: to create a new playlist and add the playing song to it
\item Next: play the next song in the album/artist/all songs/playlist queue (Alert if last song in queue)
\item Previous: play the previous song in the album/artist/all songs/playlist queue (Alert if first song in queue)
\end{itemize}

User feedback suggested that the interface exit the activity and resume playback if an appropriate response is not spoken within a given time and to provide a cancel option to exit an activity. This prototype caters to users who require a completely hands free experience by using voice as a medium to accomplish the task. The interface should include a tutorial mode for easier onboarding of novice users too.

\begin{figure}[hbt!]
\centering
\resizebox*{7cm}{7cm}{\includegraphics{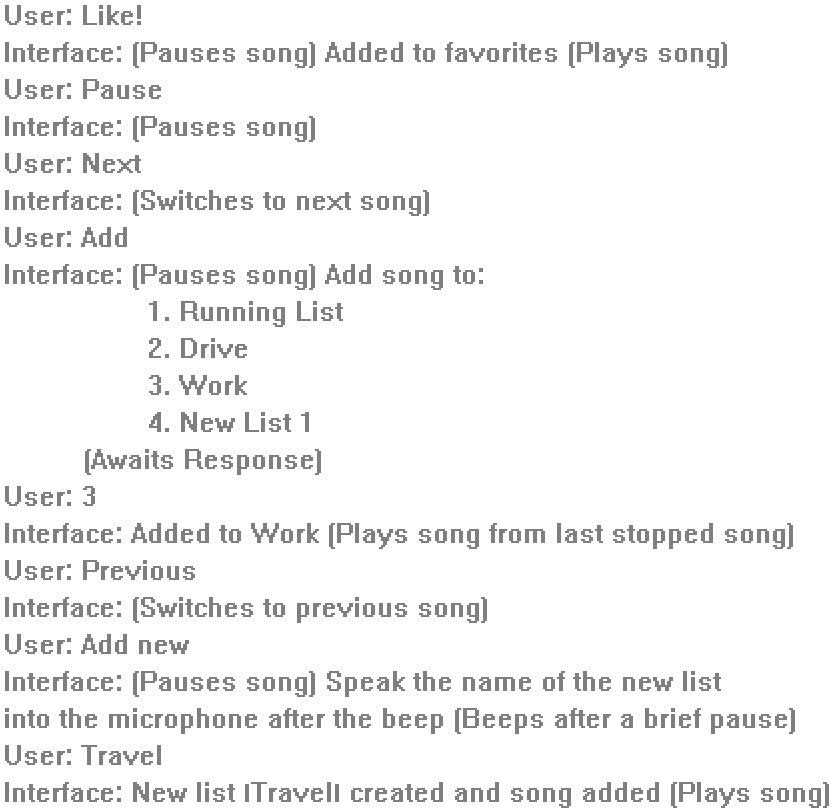}}\hspace{5pt}
\caption{Script of the 'Wizard of Oz' prototype, the song is assumed to be playing in the background.} \label{woz-prototype}
\end{figure}

\section{Evaluation}
\subsection{Quantitative Evaluation}
Participants recruited were those who used Android devices for music playback. The participants in the evaluation were a mix of novice and advanced users from varied backgrounds. The users ranged from casual music listeners to audiophiles and musicians and also people with strong technology backgrounds.
The prototype evaluated is the Wizard of Oz prototype described previously. The evaluation process was a post event protocol, where the users were provided a simulated experience of a voice based interface.
Since the prototype is a Wizard of Oz prototype for a voice controlled interface, the participants were provide with a sheet of commands and the actions they are mapped to. Data pertaining to deviation from the default commands, the keywords for such deviations, and the response to the feedback messages was noted. The participants were asked the ease of recalling the voice commands. Also the participants were asked if the feedback to commands was at an intrusive level. They were also be asked if the default commands and their mapping to actions need to be expanded to include more common parlance in vocabulary.
\subsection{Empirical Evaluation}
The empirical evaluation phase aimed to test an experimental modification to the existing interface as seen in the card prototypes. These modifications include functionality for multiple selections of music for performing actions such as adding them to a playlist. The control interface in this instance was the existing interface that lacks the functionality to make multiple selections.
The measured variable was the time taken to add 10 songs to a playlist. The null hypothesis was the assumption that the times taken to create playlists are equal regardless of the presence of the multi select functionality. The alternative hypothesis was that the multi select functionality reduces the time taken to create playlists.
\newpage
To summarize:
\begin{itemize}
\item $H_{null} : t_{M.S} = t_{S.S}$
\item $H_{alt}  : t_{M.S} < t_{S.S}$
\end{itemize}
The test was administered within subjects with the interfaces presented at random orders in order to curtail bias. The participants were chosen randomly to gather evaluation from various expertise levels. The subjects were asked to add 10 songs to a playlist in both treatments and the time taken to do so in both instances was measured. Possible lurking variables could have been fatigue arising from performing one treatment leading to poor performance in the next, or unfamiliarity with multi select interfaces leading to larger time taken.
The chi-squared test or median test was then used to evaluate the validity of the alternate hypothesis.

\section{Results}
\subsection{Post event feedback from users for Wizard of Oz prototype}
\begin{itemize}
\item User 1: Interface responses could have been a little slower in order that the user can keep up with what it is saying. Also a little confusion between the Add and Add New commands, maybe have dissimilar keywords for the two commands. Possibly have some more keywords for the ‘add to favorites’ option other than like.
\item User 2: Liked the overall experience of the interface, add to existing list might be cumbersome if there are a lot of existing lists as it might interrupt playback for a while. Also make the interface resume playback from a few seconds before it stopped.
\item User 3: Possibly have a restart option, for playing a song again in case someone zones out during a song. Also have a stop option, different from pause, which will not resume from the spot it left off when next played.
\item User 4: How will the interface deal with non-standard playlist names, when spoken into the microphone? Can it spell these accurately? Maybe include a keyboard input for such operations.
\item User 5: Listing playlist names by number to add to is a little cumbersome and reminds one of an automated telephone interface with some customer care number. Since the keypad is used for this operation eventually, just have the user select the list from the screen
\end{itemize}

\subsection{Raw Data and test results for Card Prototype}
Based on the data from Table 1. after performing the Chi-squared test, the P-Value is calculated as: 6.20746E-40 and the Probability from Student’s T-test is calculated as: 2.18931E-08 based on which interpretations the null hypothesis is rejected since it is a very low percentage that the alternate hypothesis is true due to random chance.
\begin{table}
\tbl{Raw data results for Card Prototype.}
{\begin{tabular}{lcc} \toprule
 Multiple Select Option (Prototype) & Single Select Option (Existing Interface) \\ \midrule
 46 & 113\\
 50 & 116\\
 42 & 108\\
 49 & 114\\
 53 & 117\\ \bottomrule
\end{tabular}}
\tabnote{Time taken (in seconds) to create a playlist of 10 songs with the prototype versus the existing interface}
\label{result-table}
\end{table}
\newpage
\section{Inferences}
\subsection{Wizard of Oz prototype}
The analysis of the sessions yielded the following insights:
\begin{itemize}
\item The keywords dictionary needs to be expanded, in that there is a need for synonymy.
\item The interface responses were not as intrusive to the listening experience as earlier imagined.
\end{itemize}
A major take away is that voice alone cannot act as an efficient interface. In situations such as adding to an existing playlist, it is more feasible to present the list of existing playlists on screen, rather than have a voice call it out. However, it does have its merits in that it can be an efficient playback control and can also perform rudimentary ordering and adding in situations that demand hands free usage.
\subsection{Card prototype}
Based on the data and the tests conducted, the null hypothesis can safely be rejected, since there was a significant difference in time taken to compile a playlist which could scarcely be attributed to random chance.
The results of the evaluation and analysis did match the initial expectation that a multiple selection feature would speed up the process of compiling a playlist. The results do reflect a difference in the time taken which could not be attributed to randomness or other lurking variables.
Since largely positive evaluations were received on both prototypes, I would raise the fidelity on the card prototype to a wireframe. This would help me in getting better feedback about placement of buttons, checkboxes, and other design elements. As for the Wizard of Oz prototype, I would like to investigate incorporating some visual elements into the voice interface.
\subsection{Final Thoughts}
The work conducted and the subsequent results and inferences drawn served to provide considerable clarity on the usage and need for playlists. The study primarily served to provide considerable insight into the nature of the tasks of creating, ordering,  and adding tracks to a playlist, and the effects that the interface has on the performance of these tasks. Finally, the study highlights the improvements possible by the use of common affordances based on fundamental HCI practices and possible alternative interface designs. In all, the study served to highlight the possiblities in making an interface intuitive by following basic and core HCI design principles.

\section*{Acknowledgement}
The author would like to express special acknowledgement for Dr. David Joyner and the team of \emph{CS 6750} at \emph{Georgia Tech} for their work on the course material and the course itself. Without having taken the course, it would not have been possible for me to write this work.

\section*{Disclosure statement}

The study performed for the writing of this article was undertaken during the Summer 2018 semester of \emph{CS 6570: Human-Computer Interaction} at Georgia Institute of Technology \emph{(Georgia Tech)}. Should any current student of the class come across this article, please be aware that reproduction of or significant similarity to prior work is a violation of the University Honor Code, and as such disciplinary action can be initiated against violators.

\end{document}